\documentclass[preprint]{revtex4-1}
\usepackage{graphicx}% Include figure files
\usepackage{bm}% bold math
\usepackage{bbm}
\usepackage{amssymb}
\begin{document}
\date{\today}

\title{
Soliton defects and topological $4\pi$-periodic superconductivity from an orbital magnetic field effect
in edge Josephson junctions 
}

\author{G. Tkachov}

\affiliation{
Institute of Physics, Augsburg University, 86135 Augsburg, Germany}

\begin{abstract}
Recently, much research has been dedicated to understanding topological superconductivity and Majorana zero modes 
induced by a magnetic field in hybrid proximity structures.
This paper proposes a realization of topological superconductivity in a short Josephson junction at an edge of a 2D topological insulator  
subject to a perpendicular magnetic field. 
The magnetic field effect is entirely orbital, coming from a gradient of the order parameter phase at the edge, 
which results in a soliton defect at the junction with a pair of gapless Andreev bound states.
The latter are reducible to Majorana zero modes by a unitary rotation and protected by a chiral symmetry.
Furthermore, both ground state and excitations are quasiperiodic in the magnetic flux enclosed in the junction, 
with the period equal to the double flux quantum $2\Phi_0 = h/e$. 
This behaviour follows from the gauge invariance of
the $4\pi$ - phase periodicity of the Majorana states and 
manifests itself as $2\Phi_0$ - spaced magnetic oscillations of the critical current.      
Another proposed observable is a persistent current occurring in the absence of an external phase bias.
Beside the oscillations, it shows a sign reversal prompted by the neutral Majorana zero modes. 
These findings offer the possibility to access topological superconductivity through low-field dc magnetotransport measurements.
\end{abstract}

\maketitle

\section{introduction}

Topological superconductors host zero-energy excitations at the boundary or on topological defects that are otherwise prohibited by the bulk pairing gap.
Depending on the system, such topological excitations come as Majorana edge or end states, zero modes in the vortex core or gapless Andreev levels 
in superconducting junctions \cite{Kitaev01,Nayak08,Beenakker13a,Sato17,Mizushima18,Haim18}. 
A suitable platform to realize topological superconductivity (TS) is the hybrid structures combining conventional superconductors with 
topological insulators \cite{Fu08,Tanaka09,Qi10} or semiconductor nanostructures \cite{Sau10,Alicea10,Lutchyn10,Oreg10}. 
Such structures are accessible experimentally, allowing one to test the theoretical predictions in electrical transport measurements 
\cite{Mourik12,Rokhinson12,Hart14,Nadj-Perge14,Wiedenmann16,Bocquillon17,He17}.

Particular attention has been paid to the role of the Zeeman effect in achieving TS with zero-energy Majorana states.
On the other hand, it has been noticed that gapless Majorana fermions can emerge also from the Landau quantization 
of the electron motion in a strong magnetic field \cite{Tiwari13}.  
The details of the superconducting behaviour in a magnetic field appear to be essential for low-energy excitations.
Typically, the effect of an external magnetic field consists in acting on the orbital motion, which is the cause of the ubiquitous diamagnetism 
of superconductors (e.g., the Meissner effect). The coupling to the electron spin can manifests itself 
in specific cases, such as a thin-film superconductor in a strong in-plane magnetic field, 
when the Meissner screening currents are reduced in favour of the Zeeman splitting.
Leaving aside such specific situations, 
one may inquire whether a classical (nonquantizing) magnetic field would be able to induce TS by acting solely on the orbital motion.  
That could grant access to hitherto unexplored regimes of TS, requiring neither large values of the g-factor nor strong magnetic fields and 
avoiding the suppression of the proximity-induced gap \cite{GT04,GT05,Rohlfing09,Maier12}.  

This paper demonstrates such a possibility in a hybrid setup realizing a superconducting junction at an edge of a 2D topological insulator (2DTI). 
We will see that in a short junction an external magnetic field (applied perpendicularly to the 2DTI plane) induces a topological defect 
akin to the soliton in the Jackiw-Rebbi model \cite{Jackiw76}. 
In their seminal work, Jackiw and Rebbi considered the 1D Dirac fermion coupled to a soliton and found a bound state at zero energy.
The present case is different in that the soliton hosts Andreev bound states (ABSs) whose energy levels disperse with the applied magnetic field. 
Still, there exists a straight connection to the Jackiw-Rebbi model:
We show that, up to a unitary rotation, the ABSs coincide with Majorana zero modes (MZMs) described by the Jackiw-Rebbi formalism.
In a short junction, the relation between the ABSs and MZMs is given by the simple formula
\begin{equation}
\Psi^{^{ABS}}(x,t) = U(t) \Psi^{^{MZM}}(x).
\label{ABS_MZM}
\end{equation}
Here, $\Psi^{^{ABS}}(x,t)$ and $\Psi^{^{MZM}}(x)$ are the corresponding spinors, and $U(t)$ is the time-dependent rotation matrix 
\begin{equation}
U(t) = 
\left[
\begin{array}{cc}
 e^{-i E_+(\Phi,\phi) t/\hbar} & 0 \\
 0 & e^{-i E_-(\Phi,\phi) t/\hbar}
\end{array}
\right].
\label{U_diag}
\end{equation}
As explained in detail below, the transformation (\ref{U_diag}) is diagonal in the basis of the chirality operator of the Jackiw-Rebbi model \cite{Jackiw76}, 
with $E_\pm(\Phi,\phi)$ being the ABS energy levels for the chirality eigenvalues $\pm 1$.
Above, $\Phi$ is the magnetic flux through an effective junction area, and $\phi$ is an external phase bias.

%%%  Figure   %%%
\begin{figure}[t]
\includegraphics[width=100mm]{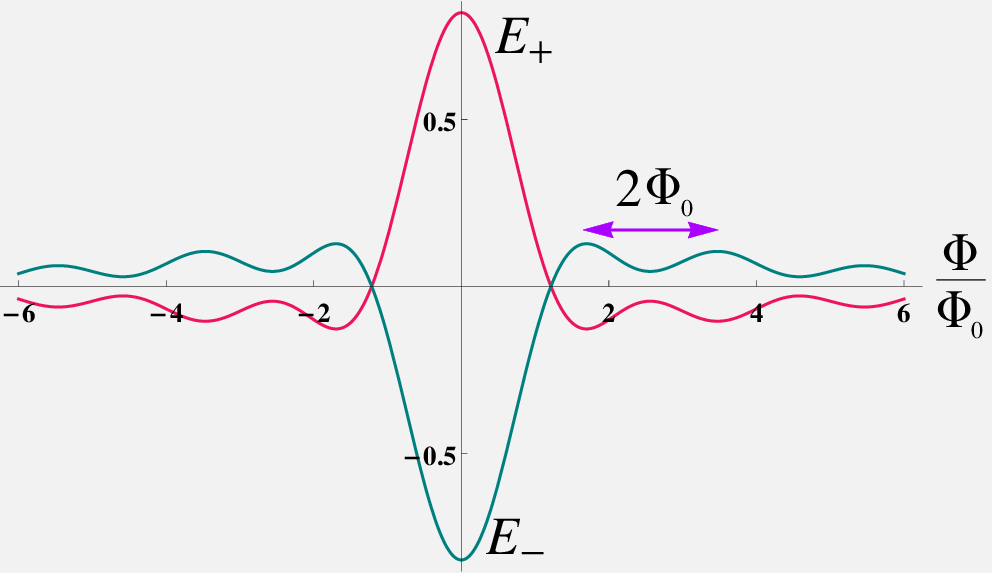}
\caption{
ABS levels $E_\pm(\Phi,0)$ (\ref{E_pm}) as function of the magnetic flux, $\Phi$, through an effective area of a superconducting edge junction.
The level crossing (protected by the chiral symmetry) as well as the $2\Phi_0$ - spaced magnetic oscillations 
indicate the magnetic-field-induced TS. The junction parameters are $W = \sqrt{\varkappa_+ \varkappa_-}w = 4$ 
and $\varkappa_- / \varkappa_+ = 0.25$ (see also text).   
}
\label{Fig_E}
\end{figure}

To briefly announce the main findings: 
{\bf i}) In a perpendicular magnetic field, the opposite-chirality ABS levels cross at zero energy, realizing a pair of the MZMs 
in the absence of the external phase bias (for $\phi = 0$, see also Fig. \ref{Fig_E}), 
{\bf ii}) the level crossing is accompanied by magnetic oscillations with the single-electron flux spacing $2\Phi_0 =h/e$, and 
{\bf iii}) these spectral features of the induced TS are observable in the equilibrium dc Josephson transport.

We note that the $2\Phi_0$-spaced oscillations in the magnetic flux translate into a $4\pi$ periodicity in the Josephson phase difference. 
The $4\pi$ - periodic Josephson effect has been put forward as a signature of MZMs in model $p$-wave superconductors \cite{Kitaev01,Kwon04}, 
which has caused a surge of interest in the related phenomena in the hybrid structures, both in theory 
(see, e.g., Refs. \cite{Fu09,Badiane11,Dominguez12,San-Jose12,Pikulin12,GT13,Beenakker13b,Badiane13}) and 
in experiment \cite{Rokhinson12,Wiedenmann16,Laroche17}.  
Most of the research on the $4\pi$ Josephson effect has been dealing with the dynamics of superconducting junctions under external driving. 
This paper suggests a different approach based on the action of the perpendicular magnetic field on the superconducting edge states.
It is essential that in real space the edge states are two-dimensional, spreading exponentially into the 2DTI bulk. 
This enables a nonlocal proximity-induced pairing that depends on the magnetic flux in the junction, $\Phi$,
thus making the topological $4\pi$ periodicity observable through the $2\Phi_0$ magnetic oscillations.
The following sections explain the details of the calculations and provide an extended discussion of the results.

\section{Superconducting edge junction in a perpendicular magnetic field: Model}
\label{Model}

%%%  Figure   %%%
\begin{figure}[t]
\begin{center}
\includegraphics[width=100mm]{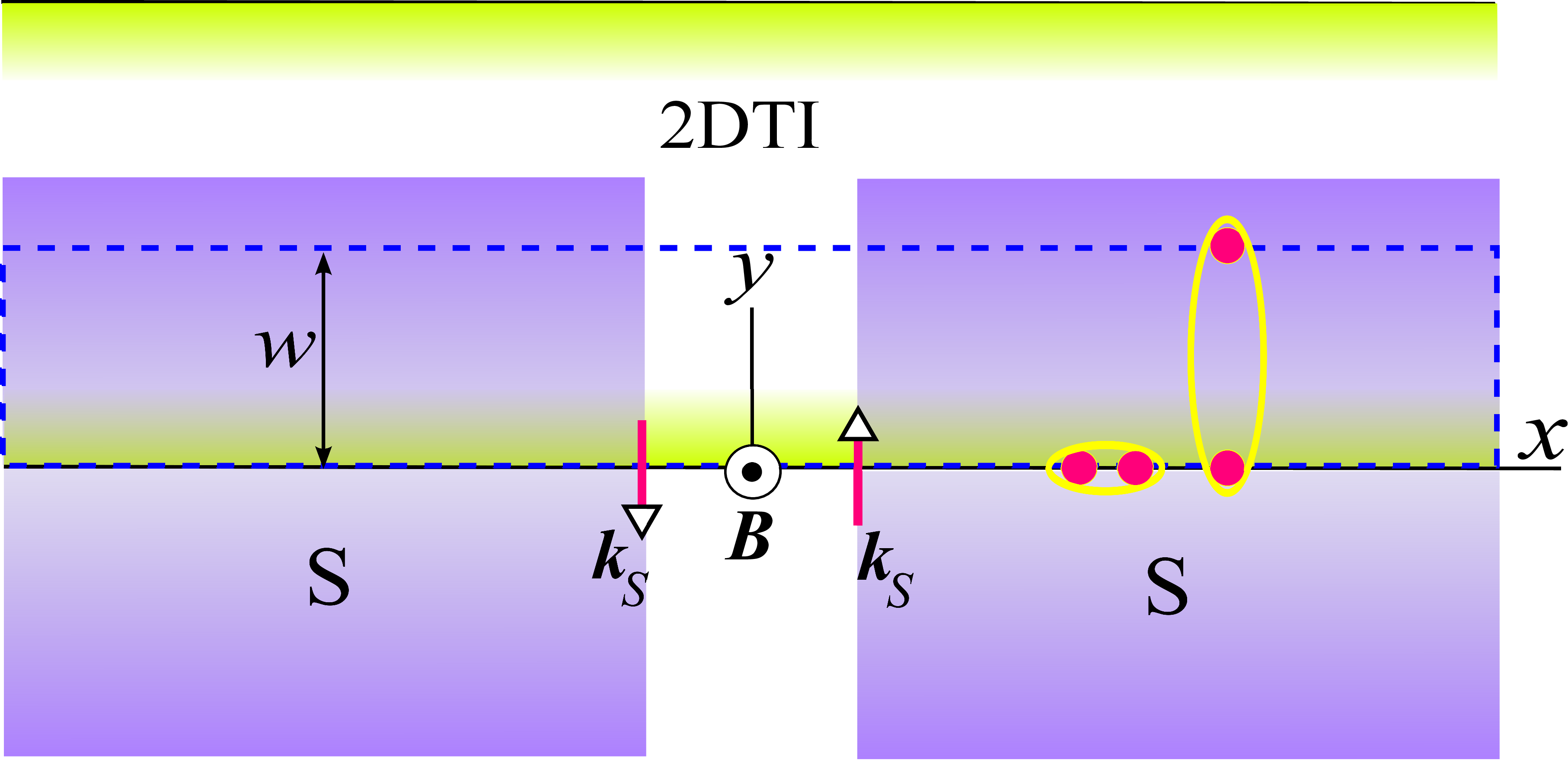}
\end{center}
\caption{
Superconducting junction at an edge of a 2DTI in a perpendicular magnetic field ${\bm B}$. 
A twist in the superconducting phase gradient, ${\bm k}_{_S}$, across the junction 
produces a soliton-like topological defect hosting MZMs. 
The spreading of the edge state into the 2DTI bulk results in the dependence of 
the Josephson transport on the magnetic flux enclosed in the effective junction area 
(indicated by the dashed contour). $w$ is the 2DTI half-width, $L$ is the length of the superconducting lead.    
}
\label{S_edge}
\end{figure}

We consider a junction between two conventional superconductors placed on one of the edges of a 2DTI, 
as sketched in Fig. \ref{S_edge}. The superconductors cover completely a half of the 2DTI width, $w$, without contacting the other edge. 
The proximitized regions play the role of the superconducting leads, 
each described by the Bogoliubov-de Gennes (BdG) Hamiltonian (see also Appendix \ref{Appendix_BdG}):
\begin{eqnarray}
{\cal H} = 
\left[%
\begin{array}{cc}
\upsilon s_z p_x - \mu & \Delta(x) \\
 \Delta^*(x)  & -(\upsilon s_zp_x - \mu)
\end{array}
\right],
\,\, 
\upsilon = \frac{|{\cal A}|}{\hbar}{\rm sgn}({\cal M}).
\label{H}
\end{eqnarray}
Here, $\upsilon$ is the edge-state velocity, 
$p_x=-i\hbar\partial_x$ is the momentum operator, $s_z$ is the spin Pauli matrix, and 
\begin{eqnarray}
\Delta(x) = \frac{\int^w_0 \Delta_{_{2D}}(x,y) f(y)dy}{\int^w_0 f(y)dy} 
\label{Delta}
\end{eqnarray}
is the effective edge pair potential. 
It is obtained by averaging the 2D proximity-induced pair potential, $\Delta_{_{2D}}(x,y)$, 
over the edge-state profile described by the transverse wave function 
\begin{equation}
f(y)= e^{ -\varkappa_+ y } - e^{ -\varkappa_- y }, \quad
\varkappa_\pm=\frac{|{\cal A}|}{2|{\cal B}|} \pm \sqrt{ \frac{{\cal A}^2}{4{\cal B}^2} + \frac{ {\cal M} }{ {\cal B} } },
\label{f}
\end{equation}
where $\varkappa_\pm$ are the decay constants depending on the band structure parameters ${\cal A}, {\cal B}$, and ${\cal M}$ 
(see Refs. \cite{Bernevig06,Liu16}). The edge states exist for an inverted band structure when
\begin{equation}
\varkappa_+\varkappa_-= - {\cal M}/{\cal B} > 0. 
\label{inverted}
\end{equation}
In Eq. (\ref{Delta}), the integration limits are determined by the boundary conditions 
at the edge ($y=0$) and in the middle ($y=w$) of the 2DTI (see also Appendix \ref{Appendix_BdG}).
The 2DTI half-width $w$ is thus an effective width of the junction. 

The effect of a weak (nonquantizing) perpendicular magnetic field can be accounted for by a local phase $\varphi(x,y)$ of 
the proximity-induced pair potential \cite{GT04,GT05,Rohlfing09,Maier12,Kopnin05,GT17}:

\begin{eqnarray}
\Delta_{_{2D}}(x,y)= \Delta_0 e^{ i\varphi(x,y) }, \,  
\varphi(x,y) =
\left\{
\begin{array}{c}
\varphi_0 + k_{_S}y \,\, {\rm (right)}, \\
{\bar \varphi}_0 - k_{_S}y \,\, {\rm (left)},
\end{array}
\right.
\label{Delta_2D}
\end{eqnarray}
where $\Delta_0$ is a real constant. The magnetic field 
generates a phase gradient, $\pm k_{_S}$, in the transverse ($y$) direction, 
with the opposite signs in the right and left leads.   
The phase is counted from its value at the edge which is
$\varphi_0$ (${\bar \varphi}_0$) on the right (left).
The relation between $k_{_S}$ and the magnetic field can be obtained from gauge-invariant Stokes' formula 
$
\oint \nabla \varphi \cdot d{\bm l} = 2\pi \Phi_{_S}/\Phi_0,
$
where the integration path runs along the boundary of a superconductor, closing through its interior  
such that there $\nabla \varphi = 0$ (see dashed path in Fig. \ref{S_edge} and \cite{Geometry}).  
Then, the phase gradient at the boundary is proportional to the magnetic flux 
through the enclosed area in the superconductor: $k_{_S} = 2\pi \Phi_{_S}/(w\Phi_0)$. 
The total flux in the junction, $\Phi = 2\Phi_{_S} + \Phi_{_N}$, includes the contributions of the leads and that of the normal region between them, 
$\Phi_{_N}$. Technically, for lateral proximity structures it is very common that $\Phi_{_N} < 2\Phi_{_S}$ 
(see, e.g., Refs. \cite{Rohlfing09,Maier12,Oostinga13}). We can therefore assume a short junction where $\Phi_{_N} \ll 2\Phi_{_S}$ and 
\begin{equation}
k_{_S} \approx \frac{\pi}{w} \frac{\Phi}{\Phi_0}.
\label{k_S}
\end{equation}
From Eqs. (\ref{Delta}) - (\ref{Delta_2D}), we find the effective edge pair potential (e.g., in the right lead) as
\begin{eqnarray}
\Delta = \bigg[
\Delta_1 e^{ i\vartheta_1 } + \Delta_2 e^{ i(\vartheta_2 + k_{_S}w)}
\bigg]
e^{ i\varphi_0 }, 
\label{Delta_sum}
\end{eqnarray}
where 
\begin{eqnarray}
\Delta_1 &=&  \Delta_0 \frac
{\varkappa_+\varkappa_-}
{\sqrt{
(\varkappa_+\varkappa_- - k_{_S}^2)^2 + (\varkappa_+ + \varkappa_-)^2 k_{_S}^2
}
},
\label{Delta1}\\
\vartheta_1 &=& \arctan \frac{ (\varkappa_+ + \varkappa_-)k_{_S} }{\varkappa_+\varkappa_- - k_{_S}^2},
\label{theta1}\\
\Delta_2 &=& \Delta_0
\frac
{ \varkappa_+\varkappa_- e^{-\varkappa_- w}}
{(\varkappa_- - \varkappa_+)
\sqrt{
\varkappa^2_- + k_{_S}^2
}
},
\label{Delta2}\\
\vartheta_2 &=& \arctan \frac{ k_{_S} }{\varkappa_-}.
\label{theta2}
\end{eqnarray}
In Eq. (\ref{Delta_sum}), the first term is the local pairing, with both electrons being at the edge $y=0$, 
while the second term accounts for nonlocal pairs, with one electron at the edge and the other at a distance of order of $w$ 
(see also Fig. \ref{S_edge}). The remote electron picks up a phase $k_{_S}w$ related to the magnetic flux in the junction. 
Above, the 2DTI half-width $w$ is assumed large compared to the edge-state width, 
so it suffices to keep only the exponential factor $e^{-\varkappa_- w}$ with the smallest decay constant $\varkappa_-$ 
in the nonlocal pairing amplitude $\Delta_2$ [see Eq. (\ref{Delta2})]. 
Finally, the phases $\vartheta_1$ and $\vartheta_2$ in Eq. (\ref{Delta_sum}) originate from the 
spatial oscillations of the order parameter on the scale of the edge-state width.
Apart from the broken time reversal, the generation of the intrinsic phases $\vartheta_1, \vartheta_2$, and $k_{_S}w$ 
requires an inversion asymmetry under $y \to -y$, which is encoded in the edge wave function (\ref{f}).

\begin{figure}[t]
\begin{center}
\includegraphics[width=100mm]{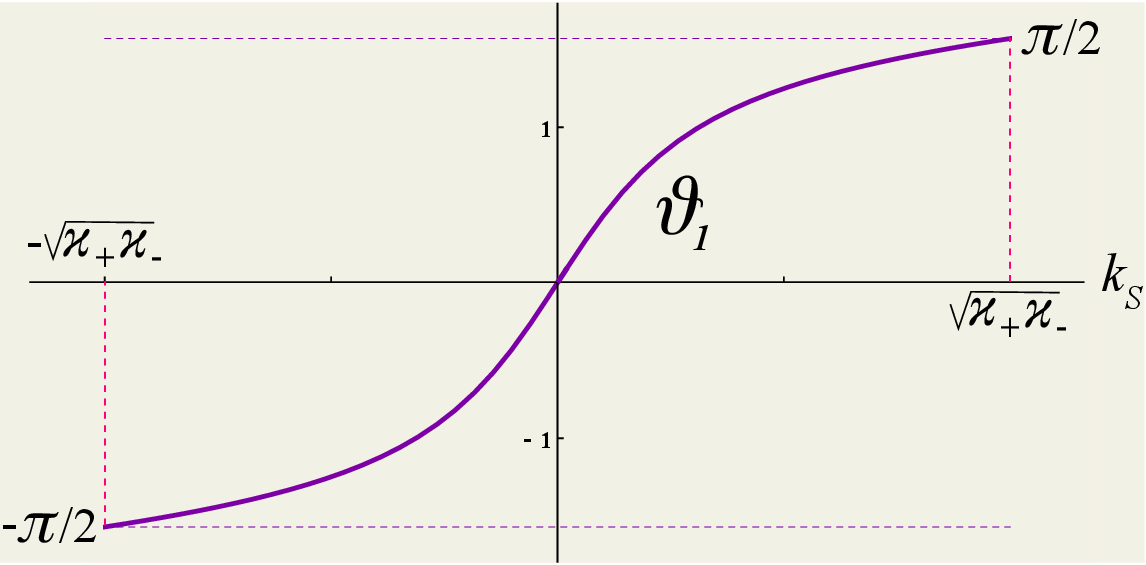}
\end{center}
\caption{
The dependence of $\vartheta_1$ (\ref{theta1}) on the superconducting phase gradient $k_{_S}$,
indicating a soliton defect at the junction; $\varkappa_- / \varkappa_+ = 0.25$.
}
\label{theta1_fig}
\end{figure}

Noteworthy is the behaviour of $\vartheta_1$ [see Fig. \ref{theta1_fig} and Eq. (\ref{theta1})].
It takes the universal values $\pm \pi/2$ as the phase gradient $k_{_S}$ reaches $\pm \sqrt{\varkappa_+\varkappa_-}$,
implying a $\pi$ phase drop across the junction in the absence of any external phase bias.
For any $k_{_S}$, there is a phase twist $\vartheta_1 {\rm sgn}(x)$ across the junction, 
which is {\em trully topological} because it occurs only for the inverted band structure. 
The normal band ordering (${\cal M}/{\cal B} > 0$) would give an imaginary $\sqrt{\varkappa_+\varkappa_-}$, 
invalidating the model. In the following, we use the BdG Hamiltonian (\ref{H}) in the basis of the Nambu-Pauli matrices 
$\tau_1, \tau_2$, and $\tau_3$:  
\begin{eqnarray}
{\cal H}(x) = \tau_3 \upsilon s p_x + \tau_1 \Delta_{_{Re}}(x) - \tau_2 \Delta_{_{Im}}(x) ,
\label{H_Nambu}
\end{eqnarray}
where $\Delta_{_{Re}}(x)$ and $\Delta_{_{Im}}(x)$ are the shorthand notations for the real and imaginary parts of the junction pair potential
\begin{equation}
\Delta(x) = \Delta_1 e^{i(\vartheta_1 + \phi/2) {\rm sgn}(x)} 
+
\Delta_2 e^{i(\vartheta_2 + \phi/2 + \pi\Phi/\Phi_0 ){\rm sgn}(x)},
\label{Delta_sum1}
\end{equation}
and $\phi = \varphi_0 - {\bar \varphi}_0$ is the external phase bias 
[the average phase $(\varphi_0 + {\bar \varphi}_0)/2$ and the chemical potential have both been gauged out]; 
$s$ is an eigenvalue of $s_z$. 

%%%  Figure   %%%

\section{Emergent soliton, Majorana zero modes and relation to Andreev bound states}
\label{ABSs}

Next, we turn to the fermionic excitations in a short junction described by the BdG equation $i\hbar\partial_t \Psi(x,t) = {\cal H}(x) \Psi(x,t)$. 
It is instructive to map the problem to the Jackiw-Rebbi model \cite{Jackiw76}. 
In view of the symmetry ${\cal H}^*(-x)={\cal H}(x)$, it suffices to consider $x>0$, 
where the pair potential $\Delta$ is a constant given by Eq. (\ref{Delta_sum1}). 

A unitary transformation 
\begin{equation}
\Psi(x,t) = U(t)  \, \Psi^\prime(x,t), \,\,\, {\rm with} \,\,\, U(t) = e^{-i \tau_1 \Delta_{_{Re}} t/\hbar},
\label{U}
\end{equation}
brings the BdG Hamiltonian to the form
\begin{eqnarray}
{\cal H}^\prime(x,t) = U^\dagger(t)  [\tau_3 \upsilon s p_x - \tau_2 \Delta_{_{Im}}] U(t).
\label{H^prime}
\end{eqnarray}
Albeit time dependent, it shares two essential features of the Jackiw-Rebbi model: 
It has the chiral symmetry $\tau_1 {\cal H}^\prime(x,t) \tau_1 = - {\cal H}^\prime(x,t)$, 
and the imaginary part of the pair potential $\Delta_{_{Im}}$ (odd in the full space) acts as a soliton defect. 
An eigenstate of the chirality operator $\tau_1$ 
satisfies the equation ${\cal H}^\prime(x,t) \Psi^\prime(x,t)=0$ or, 
explicitly, $(\hbar\upsilon\partial_x - s \tau \Delta_{_{Im}})\Psi^\prime(x,t)=0$, 
where $\tau = \pm 1$ is an eigenvalue of $\tau_1$. 
The solution is an MZM of the form $\Psi^\prime(x) \propto e^{-kx}$, 
where the inverse decay length is given by $k = - s\tau \Delta_{_{Im}}/(\hbar \upsilon)$, 
yielding a normalizable MZM with the spin projection
\begin{equation}
s = - \tau {\rm sgn}(\upsilon \Delta_{_{Im}}) = - \tau {\rm sgn}({\cal M} \Delta_{_{Im}}).
\label{Sign}
\end{equation}
The full MZM spinor is 
\begin{eqnarray}
\Psi^{^{MZM}}_\tau(x) = C e^{- \left|\frac{\Delta_{_{I}}(\Phi, \phi) }{\hbar\upsilon}\right| x } 
(1 + \tau \tau_1) \otimes (1 - \tau {\rm sgn}[{\cal M} \Delta_{_{Im}}(\Phi, \phi)]s_z)
\left[
\begin{array}{c}
1 \\ 0 \\ 0 \\ 0
\end{array}
\right] ,
\label{MZMs}
\end{eqnarray}
where $\otimes$ means the direct product of the Nambu and spin matrices, $C$ is the normalization constant, and 
\begin{eqnarray}
\Delta_{_{Im}}(\Phi, \phi) =
\Delta_1 \sin\left(\vartheta_1 + \frac{\phi}{2} \right) 
+ 
\Delta_2 \sin\left(\vartheta_2 + \frac{\phi}{2} + \pi\frac{\Phi}{\Phi_0} \right).
\label{Im_Delta}
\end{eqnarray}
Unlike the Jackiw-Rebbi model, the 2DTI supports an MZM for each value of $\tau = \pm 1$.
Nevertheless, the generic chiral symmetry of the BdG Hamiltonian (\ref{H^prime}) 
prevents the mixing of the MZMs, enabling access to their specific properties associated with the non-Abelian statistics \cite{Kitaev01,Fu09}.
Consider a phase translation $\phi \to \phi + 2\pi$. 
It flips the sign of the decay constant $k$,
so the state $\Psi^{^{MZM}}_\tau(0)$ switches to the opposite side of the region with the pairing gap $\Delta$,
and another phase advance of $2\pi$ is needed to bring this state back \cite{Kitaev01,Fu09}.
As seen from Eqs. (\ref{MZMs}) and (\ref{Im_Delta}), 
the MZMs are indeed $4\pi$ - periodic in the Josephson phase difference $\phi$ 
and, by gauge invariance, $2\Phi_0$ - quasiperiodic in the magnetic flux $\Phi$ enclosed in the junction.

In the chosen representation (\ref{U}), 
the ABS spinors are obtained by a unitary time-dependent rotation $U(t)$ of the MZM spinors (\ref{MZMs}).
Furthermore, since $U(t)$ and $\tau_1$ commute, the ABSs appear to be the eigenstates of the chirality as well,
thus inheriting this key property of the MZMs. 
In the chirality basis, the rotation matrix $U(t)$ (\ref{U}) becomes diagonal, 
so the final result for the ABSs is rather simple and given by Eqs. (\ref{ABS_MZM}) and (\ref{U_diag}) from the introduction. 
Comparing Eqs. (\ref{U_diag}) and (\ref{U}), we see that the spectrum of the matrix $U(t)$ consists of 
two eigenvalues $e^{- iE_\pm(\Phi,\phi)t/\hbar}$, where the energies $E_\pm(\Phi,\phi)$ are given by $E_\pm(\Phi,\phi) = \pm \Delta_{_{Re}}(\Phi,\phi)$.
These are the ABS levels of a short Josephson junction. Taking the real part of $\Delta$ in Eq. (\ref{Delta_sum1}), we have explicitly
\begin{eqnarray}
E_\pm(\Phi,\phi) = \pm 
\biggl[
\Delta_1 \cos\left(\vartheta_1 + \frac{\phi}{2} \right) 
+ 
\Delta_2  \cos\left(\vartheta_2 + \frac{\phi}{2} + \pi \frac{\Phi}{\Phi_0} \right)
\biggr].
\label{E_pm}
\end{eqnarray}
For a zero magnetic field $B$, the ABS levels $E_\pm(0,\phi)$ (\ref{E_pm}) oscillate with the Josephson phase difference $\phi$, 
crossing periodically in the middle of the gap. A similar behaviour would be expected for any gapless ABSs. 
It can be observed in a SQUID setup where the Josephson junction is inserted into a superconducting ring with a magnetic flux $\Phi_{ring}$ 
inducing the phase drop $\phi = 2\pi (\Phi_{ring}/\Phi_0)$ across the junction.   

The orbital magnetic-field effect discussed in this paper offers a conceptually different possibility to control the ABSs.   
It does not require any external phase bias, as even for $\phi=0$
there is a soliton defect from the topological phase twist $\vartheta_1 {\rm sgn}(x)$ 
which binds the fermions at the junction. Instead of the ring flux $\Phi_{ring}$,  
the magnetic flux enclosed in the junction, $\Phi$, drives the midgap level crossing 
protected by the chiral symmetry of the BdG Hamiltonian (see also Fig. \ref{Fig_E}).  
The MZMs at the level crossing and the $2\Phi_0$ magnetic oscillations are the hallmarks of the magnetic-field-induced TS. 
It is the nonlocality of the pairing that makes the correspondence between the $2\Phi_0$ magnetic oscillations and the topological $4\pi$ - phase periodicity 
apparent, see the second term in Eq. (\ref{E_pm}) which depends on the gauge-invariant total phase difference $\phi + 2\pi \Phi/\Phi_0$.
 
\section{Equilibrium dc Josephson transport}

This section addresses the observability of the magnetic-field-induced TS 
in equilibrium dc Josephson transport. We begin by calculating the Josephson current-phase relationship \cite{Golubov04}: 
$J(\phi) = (2e/\hbar) (\partial E_+/\partial\phi) [n(E_+) - 1/2]$,
where $n(E_+)$ is the occupation of the ABS level $E_+$ [see Eq. (\ref{E_pm})].
At zero temperature and in equilibrium, $n(E_+) = [1 - {\rm sgn}(E_+)]/2$. 
Hence, $J(\phi) = -(e/\hbar) (\partial E_+/\partial\phi) {\rm sgn}(E_+)$ and, finally, 

\begin{eqnarray}
J(\Phi, \phi) =  
\frac{e}{2\hbar} 
{\rm sgn} 
& \Biggl[ &
\Delta_1 \cos\left(\vartheta_1 + \frac{\phi}{2}\right) + \Delta_2 \cos\left(\vartheta_2 + \frac{\phi}{2} + \pi\frac{\Phi}{\Phi_0} \right)
\Biggr] 
\nonumber\\
\times
& \Biggl[ &
\Delta_1 \sin\left(\vartheta_1 + \frac{\phi}{2} \right) + \Delta_2 \sin\left(\vartheta_2 + \frac{\phi}{2} + \pi\frac{\Phi}{\Phi_0} \right)
\Biggr] .
\label{J}
\end{eqnarray}
The subscripts 1 and 2 correspond to the local and nonlocal transport channels, reflecting the structure of the pair potential 
[cf. Eqs. (\ref{Delta_sum}) -- (\ref{Delta_sum1})]. 
We note that the switching of the level occupation upon the phase advance $\phi \to \phi + 2\pi$ 
makes the current-phase relationship $2\pi$ periodic.
However, the induced TS can still be identified through the $2\Phi_0$ - spaced magnetic oscillations in the flux $\Phi$. 
This becomes possible because the flux modulation of the nonlocal contribution does not alter the level occupation 
which is fixed by the total ABS energy. A suitable observable is the critical current $J_c(\Phi)$ defined as 
the maximum value of $J(\Phi, \phi)$ in the phase interval $0 \leq \phi \leq 2\pi$ for given flux value $\Phi$.
From Eq. (\ref{J}) we readily find

%%%  Figure  %%%
\begin{figure}[t]
\includegraphics[width=100mm]{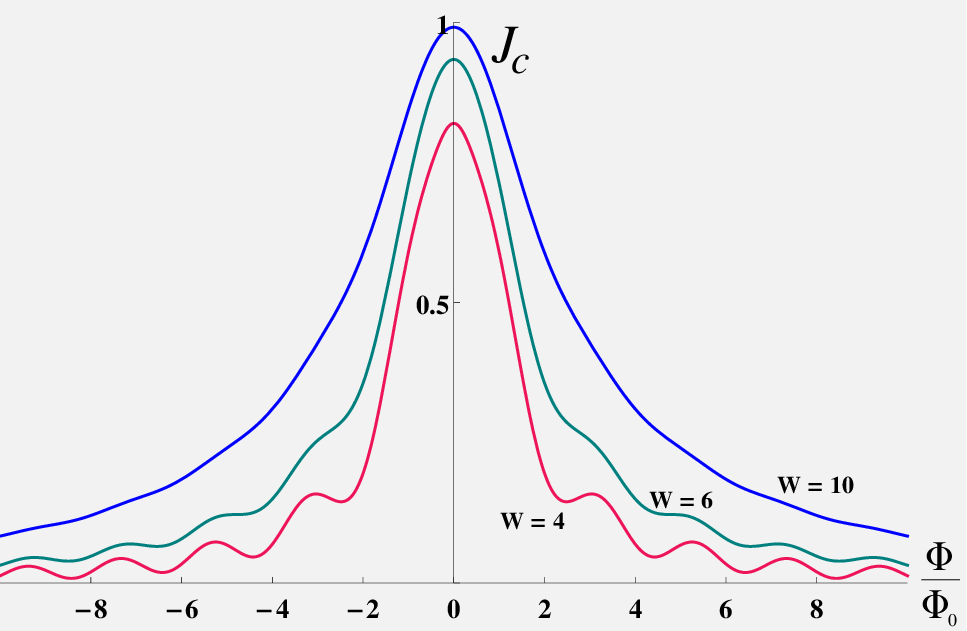}
\caption{
Critical current $J_c(\Phi)$ (\ref{J_c}) for different values of dimensionless junction width $W = \sqrt{\varkappa_+ \varkappa_-}w$ 
and $\varkappa_- / \varkappa_+ = 0.25$.
}
\label{Fig_Jc}
\end{figure}

\begin{eqnarray}
J_c(\Phi) = \frac{e}{2\hbar} 
\left[
\Delta_1 + \Delta_2 \cos\left( \vartheta_2 -  \vartheta_1 + \pi\frac{\Phi}{\Phi_0}  \right)
\right],
\label{J_c}
\end{eqnarray}
where $\vartheta_2 -  \vartheta_1 + \pi\Phi/\Phi_0$ is the relative phase difference between the nonlocal and local transport channels.  
The critical current (\ref{J_c}) decreases with the magnetic flux, showing the $2\Phi_0$ - spaced oscillations on top of 
the monotonic downturn (see Fig. \ref{Fig_Jc}). The oscillations are clearly visible if the 2DTI thickness is not too large compared to the edge-state width.

\begin{figure}[t]
\includegraphics[width=100mm]{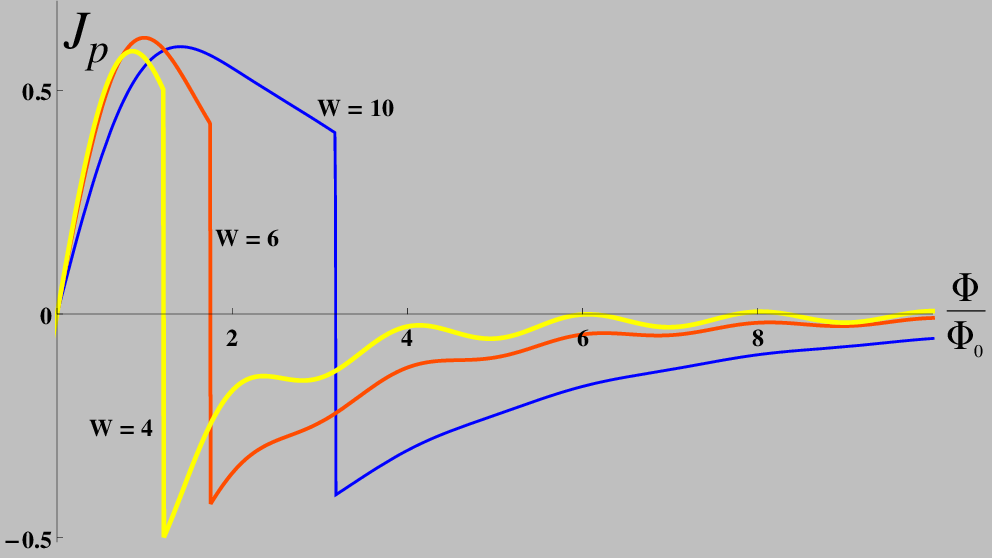}
\caption{
Persistent current $J_p(\Phi)$ (\ref{J_p}) for different values of dimensionless junction width $W = \sqrt{\varkappa_+ \varkappa_-}w$ 
and $\varkappa_- / \varkappa_+ = 0.25$.
}
\label{Fig_Jp}
\end{figure}

Another manifestation of the magnetic-field-induced TS is the persistence of the dissipationless electric current (\ref{J})
after the external phase bias is switched off. Using the notation $J_p(\Phi) = J(\Phi, 0)$ for the persistent current, we have 

\begin{eqnarray}
J_p(\Phi) = \frac{e}{2\hbar} 
{\rm sgn}
& \Biggl[ &
\Delta_1 \cos\vartheta_1 + \Delta_2 \cos\left(\vartheta_2 + \pi\frac{\Phi}{\Phi_0}  \right)
\Biggr] 
\nonumber\\
\times
&\Biggl[&
\Delta_1 \sin\vartheta_1 + \Delta_2 \sin\left( \vartheta_2 + \pi\frac{\Phi}{\Phi_0}  \right)
\Biggr].
\label{J_p}
\end{eqnarray}
Beside the $2\Phi_0$ oscillations, the persistent current (\ref{J_p}) allows the detection of the ABS level crossing and the MZMs.  
Figure \ref{Fig_Jp} shows that the function $J_p(\Phi)$ reverses the sign, which occurs at the same field as the level crossing 
(cf. plots for $W=4$ in Figs. \ref{Fig_E} and \ref{Fig_Jp}).
Furthermore, an abrupt reversal of $J_p(\Phi)$ is a signature of the {\rm linear} field dependence of the levels at the crossing point.
The fact that the persistent current goes through zero in a finite magnetic field indicates the electrically neutral MZMs. 

%%%  Figure   %%%

\section{Estimates and discussion}
\label{Estimates}

Let us estimate the magnetic field needed to localize the MZMs and discuss possible experimental setups to measure 
the currents (\ref{J_c}) and (\ref{J_p}). The characteristic magnetic field $B_{_{MZM}}$ corresponds to a $\pi$ phase drop at the junction,
which holds under condition $k_{_S} \approx (\varkappa_+\varkappa_-)^{1/2}$, where $k_{_S} = \pi B_{_{MZM}} a /(w\Phi_0)$ 
is proportional to the effective area of the junction, $a$ [cf. Eq. (\ref{k_S})].   
For lateral proximity structures (see, e.g., Refs. \cite{Rohlfing09,Maier12,Oostinga13}), 
the area $a = L w$ scales with the length, $L$, of the superconductors placed on top of the normal system.
Therefore, the characteristic magnetic field is given by 

\begin{equation}
B_{_{MZM}} \approx \frac{\Phi_0}{\pi L (\varkappa_+\varkappa_-)^{-1/2}},
\label{B_MZM}  
\end{equation}
where $L (\varkappa_+\varkappa_-)^{-1/2}$ is the effective area occupied by the edge state in the junction.  
In other words, $B_{_{MZM}}$ is the field one needs to apply in order to insert a flux quantum into the edge-state area. 
For the typical band-structure parameters of inverted HgTe quantum wells (${\cal M} \approx -10$ meV and ${\cal B} \approx 1000$ meV$\cdot$nm), 
the edge-state spreading is $(\varkappa_+\varkappa_-)^{-1/2} \approx 10$ nm [cf. Eq. (\ref{inverted})].
The junction length $L$, however, is of order of a micron. For $L = 1 \mu$m we find 
\begin{equation}
B_{_{MZM}} \approx 0.1 T.
\label{B_MZM_1}  
\end{equation}
Although the effective junction length $L$ may be somewhat smaller, 
the typical numbers for $B_{_{MZM}}$ still make up a small fraction of the Tesla.
In this case, the inclusion of the Zeeman coupling with typical g-factors ($\sim 1 - 10$) 
would neither spoil nor significantly modify the results.
Both features of the magnetic-field-induced TS, viz. the MZMs and the $2\Phi_0$ - quasiperiodic oscillations, 
are expected to occur in the same field range as the classical magnetotransport. 

As for the experimental realization, the calculation of the critical current (\ref{J_c}) assumes standard two-terminal measurements 
in which $J_c(\Phi)$ is inferred from the $I-V$ characteristics of a current-biased junction.
The measurement of the persistent current (\ref{J_p}) is more challenging, as it requires a ring geometry permitting a circular flow of the superconducting condensate.   
One possibility is to determine $J_p(\Phi)$ from the current-phase relationship measured by means of the rf SQUID magnetometry.       

\section{Summary}
\label{Summary}

This paper has studied a superconducting junction at an edge of a 2D topological insulator subject to a perpendicular magnetic field. 
It has been assumed that the field acts only on the orbital motion, causing a variation of the order parameter phase 
with a constant gradient in space. Unlike the previous work (e.g., Refs. \cite{GT04,GT05}), 
we choose the geometry in which the phase gradient is perpendicular to the conducting channel and   
switches the sign between the superconducting banks.
In this case, the magnetic field produces a phase twist across the junction, creating Andreev bound states in the absence of any external phase bias. 
Furthermore, we have found a unitary map between the Andreev bound states and the Majorana zero modes of the relativistic Jackiw-Rebbi model. 
This indicates the possibility of topological superconductivity controlled by the orbital magnetic field effect.
Its characteristic feature is the $2\Phi_0$-spaced oscillations with the magnetic flux enclosed in the junction, 
which are intimately related to the $4\pi$ - phase periodicity associated with the non-Abelian statistics of the Majorana zero modes. 
The paper has discussed the observability of the magnetic-field-induced topological superconductivity in the critical and persistent Josephson currents.

\acknowledgments
The author thanks U. Eckern for discussions.
This work was supported by the German Research Foundation (DFG) through TRR 80.

\appendix

\section{Effective Bogoliubov - de Gennes Hamiltonian for a 2DTI edge}
\label{Appendix_BdG}

In the following, we derive the effective edge BdG Hamiltonian used in the main text [see Eqs. (\ref{H}) and (\ref{Delta})]. 
It is assumed that, in the normal state, the 2DTI is described by the Bernevig-Hughes-Zhang (BHZ) Hamiltonian \cite{Bernevig06}: 

\begin{equation}
H_{_N} = {\cal A} (s_z\sigma_x k_x - s_0\sigma_y k_y) + [{\cal M} + {\cal B}( k^2_x + k^2_y)]s_0\sigma_z + {\cal C} + {\cal D}( k^2_x + k^2_y), 
\label{H_N}
\end{equation}
where $k_x$ and $k_y$ are the components of the 2D wave-vector operator ${\bm k}=[-i\partial_x, -i\partial_y, 0]$, 
$\sigma_x, \sigma_y$, and $\sigma_z$ are the Pauli matrices in the space of the $s$ - and $p$ - like orbitals, 
and $s_z$ is the spin Pauli matrix ($\sigma_0$ and $s_0$ are the corresponding unit matrices); 
${\cal A}, {\cal B}, {\cal C}, {\cal D}$, and ${\cal M}$ are the bulk band structure constants. 
In particular, ${\cal A}$ quantifies the strength of the (pseudo)spin - momentum locking, ${\cal B}$ and ${\cal D}$ characterize the band curvature, 
${\cal C}$ is the reference energy, and $2{\cal M}$ yields the energy gap between the conduction and valence bands. 
 
In contact with a conventional superconductor, placed on top of the 2DTI, the latter can be described at low energies by the BdG Hamiltonian 

\begin{equation}
H_{_{BdG}} = \tau_z H_{_N} + \tau_+ \Delta_{_{2D}}(x,y) + \tau_- \Delta^*_{_{2D}}(x,y), \qquad \tau_\pm = \frac{\tau_x \pm i\tau_y}{2},
\label{H_BdG}
\end{equation}
where $\Delta_{_{2D}}(x,y)$ is the proximity-induced pair potential in the 2DTI, 
$\tau_x, \tau_y$, and $\tau_z$ are the Pauli matrices acting in the Nambu (particle-hole) space. 
The Hamiltonian $H_{_{BdG}}$ acts on an 8-component function $\Psi(x,y)$, satisfying the equation 
$H_{_{BdG}}\Psi=E\Psi$ or, explicitly,

\begin{eqnarray}
&&
\tau_z\left\{
{\cal A} s_z\sigma_x k_x  + i{\cal A} s_0\sigma_y \partial_y  + s_0\sigma_z [ {\cal M} + {\cal B} ( k^2_x - \partial^2_y)] + {\cal C} + {\cal D} ( k^2_x - \partial^2_y)
\right\} \Psi(x,y)
+
\nonumber\\
&&
 + \tau_+ \Delta_{_{2D}}(x,y)\Psi(x,y) + \tau_- \Delta^*_{_{2D}}(x,y)\Psi(x,y) = E\Psi(x,y).
\label{Eq_BdG}
\end{eqnarray}
The 2DTI has the strip geometry with the width, $2w$, large enough to justify an independent treatment of the two edges.
We choose an edge at $y = 0$, assuming that it faces an ordinary insulator with $\Psi(x,y)=0$ in the half-space $y \leq 0$. 
This yields the boundary condition 
\begin{equation}
\Psi(x, 0) = 0.
\label{Boundary1}
\end{equation}
Another boundary condition is the requirement for $\Psi(x,y)$ to decay in the 2DTI interior:  
\begin{equation}
\Psi(x, y \to w) \to 0.
\label{Boundary2}
\end{equation}
The values of $y$ are limited by the 2DTI half-width $w$.
This is because, in reality, an edge state decays only up to the middle of a 2DTI ($y=w$), 
then the decay turns into an increase due to the presence of the other edge. 
The 2DTI half-width $w$ yields the physical cutoff for the transverse coordinate $y$ in our single-edge problem. 
   
We intend to derive from Eq. (\ref{Eq_BdG}) a 1D equation that depends only on the position along the edge, $x$. 
For a generic $\Delta_{_{2D}}(x,y)$, the separation of the $x$ and $y$ variables does not work. 
The idea is to integrate out the dependence on the transverse coordinate $y$. 
Integrating all terms in Eq. (\ref{Eq_BdG}) from $0$ to $w$, we have

\begin{eqnarray}
&&
\tau_z\Biggl\{
{\cal A} s_z\sigma_x k_x \int\limits^w_0 \Psi(x,y)dy + i{\cal A} s_0\sigma_y [\Psi(x,w) - \Psi(x,0)] +  
\label{Eq_BdG_int1}\\
&&
s_0\sigma_z \int\limits^w_0 [{\cal M} + {\cal B} ( k^2_x - \partial^2_y)] \Psi(x,y)dy  +  \int\limits^w_0 [{\cal C} + {\cal D} ( k^2_x - \partial^2_y)] \Psi(x,y)dy \Biggr\} +
\label{Eq_BdG_int2}\\
&&
\tau_+ \int\limits^w_0 \Delta_{_{2D}}(x,y) \Psi(x,y)dy + \tau_- \int\limits^w_0 \Delta^*_{_{2D}}(x,y) \Psi(x,y)dy 
= E \int\limits^w_0 \Psi(x,y)dy.
\label{Eq_BdG_int3}
\end{eqnarray}
The second term in Eq. (\ref{Eq_BdG_int1}) vanishes because of the boundary conditions (\ref{Boundary1}) and (\ref{Boundary2}). 
Furthermore, we require that the first term in Eq. (\ref{Eq_BdG_int2}) vanishes as well:

\begin{equation}
\int\limits^w_0[{\cal M} + {\cal B} ( k^2_x - \partial^2_y)] \Psi(x,y)dy=0 \quad \Longleftrightarrow \quad 
\int\limits^w_0( k^2_x - \partial^2_y) \Psi(x,y)dy= - \frac{{\cal M}}{{\cal B} } \int\limits^w_0 \Psi(x,y)dy.
\label{Gapless}
\end{equation}
This just means that the full band gap term vanishes globally. 
As in the BHZ model, the requirement (\ref{Gapless}) ensures the gapless character of the edge states [see also Eq. (\ref{Eq_transverse}) below]. 
Using the second relation in Eq. (\ref{Gapless}), we exclude the second-order derivatives $k^2_x - \partial^2_y$ in the remaining term in Eq. (\ref{Eq_BdG_int2}): 

\begin{eqnarray}
&&
\tau_z\left(
{\cal A} s_z\sigma_x k_x + {\cal C}  - \frac{ {\cal D} {\cal M} }{ {\cal B} } 
\right) 
\int\limits^w_0 \Psi(x,y)dy 
+ 
\label{Eq_BdG_int4}\\
&&
\tau_+ \int\limits^w_0 \Delta_{_{2D}}(x,y) \Psi(x,y)dy + \tau_- \int\limits^w_0 \Delta^*_{_{2D}}(x,y) \Psi(x,y)dy 
= E \int\limits^w_0 \Psi(x,y)dy.
\label{Eq_BdG_int5}
\end{eqnarray}
The use of the integral Eqs. (\ref{Eq_BdG_int1}) -- (\ref{Gapless}) allowed us to fully eliminate the second-order derivatives. 
We are left with a shift $-{\cal D} {\cal M}/{\cal B}$ of the bulk reference energy ${\cal C}$ only.  
The energy shift reflects the asymmetry between the bulk conduction and valence bands due to the "non-relativistic" quadratic term 
${\cal D}( k^2_x + k^2_y)$ in the BHZ Hamiltonian (\ref{H_N}). 
The constant $\mu = -{\cal C} + {\cal D}{\cal M}/{\cal B}$ has the meaning of the effective edge chemical potential.

Next, let us determine the edge-state profile in the $y$ direction.
For that purpose, we note that the integral equation (\ref{Gapless}) is equivalent to the differential equation used in the BHZ model \cite{Liu16}:
\begin{equation}
i{\cal A} \sigma_y \partial_y\Psi(x,y) + \sigma_z [{\cal M} + {\cal B} (k^2_x - \partial^2_y)]\Psi(x,y)=0.
\label{Eq_transverse}
\end{equation}
It can be cast to
\begin{equation}
\frac{{\cal A} }{{\cal B} }\partial_y\Psi(x,y) + \sigma_x \left(\frac{{\cal M}}{{\cal B} }+ k^2_x - \partial^2_y\right)\Psi(x,y)=0
\label{Eq_transverse1}
\end{equation}
and solved for small values of $k_x$ such that $k^2_x \ll |{\cal M}/{\cal B}|$, using the ansatz

\begin{equation}
\Psi(x,y) = \chi(x) e^{- \varkappa y}, \qquad \sigma_x \chi(x) = \eta \chi(x),
\label{Ansatz}
\end{equation}
where $\varkappa$ is the inverse decay length, and $\chi(x)$ is chosen to be an eigenstate of $\sigma_x$ with an eigenvalue $\eta$. 
Inserting Eq. (\ref{Ansatz}) into Eq. (\ref{Eq_transverse1}) yields

\begin{equation}
\varkappa^2 + \eta \frac{{\cal A} }{{\cal B} }\varkappa - \frac{{\cal M}}{{\cal B} } = 0,
\label{Eq_kappa}
\end{equation}
with the roots 
\begin{equation}
\varkappa^\eta_\pm = -\eta\frac{{\cal A} }{2{\cal B} } \pm \sqrt{\left(\frac{{\cal A} }{2{\cal B} }\right)^2 + \frac{{\cal M}}{{\cal B} }}.
\label{kappas}
\end{equation}
The normalizable edge solutions exist for an inverted band structure when $\varkappa^\eta_+\varkappa^\eta_- = - {\cal M}/{\cal B} >0$.
Moreover, in the chosen geometry, $\varkappa^\eta_+$ and $\varkappa^\eta_-$ must be both positive, 
which determines the choice of the eigenvalue $\eta$:
\begin{equation}
\eta = -{\rm sgn}({\cal A} {\cal B}) = {\rm sgn}({\cal A} {\cal M}),
\label{eta}
\end{equation}
where we use ${\rm sgn}({\cal B}) = -{\rm sgn}({\cal M})$. 
Finally, we recover the edge solution of Ref. \cite{Liu16}:

\begin{equation}
\Psi(x,y) = \chi(x) f(y), 
\quad\quad 
f(y)= e^{- \varkappa_+ y} - e^{- \varkappa_- y}, 
\quad 
\chi(x) = [1 + {\rm sgn}({\cal A} {\cal M})\sigma_x]
\left[ 
\begin{array}{c}
\Psi(x) \\ 0
\end{array}
\right],
\label{Solution}
\end{equation}
where 
$\varkappa_\pm \equiv \varkappa^\eta_\pm \bigl|_{\eta=-{\rm sgn}({\cal A} {\cal B})} 
= \frac{|{\cal A}|}{2|{\cal B}|} \pm \sqrt{ \frac{{\cal A}^2}{4{\cal B}^2} + \frac{{\cal M}}{ {\cal B} }}$,
and $\Psi(x)$ is a 4-component function (in the spin and Nambu spaces).

Inserting Eq. (\ref{Solution}) into Eqs. (\ref{Eq_BdG_int4}) and (\ref{Eq_BdG_int5}), we arrive at the 1D equation for $\Psi(x)$:

\begin{equation}
\left[
\tau_z
\left(
|{\cal A}|{\rm sgn}({\cal M}) s_z k_x - \mu 
\right)  
+ 
\tau_+ \frac{\int\limits^w_0 \Delta_{_{2D}}(x,y)f(y)dy}{\int\limits^w_0 f(y)dy} 
+ \tau_- \frac{\int\limits^w_0 \Delta^*_{_{2D}}(x,y)f(y)dy}{\int\limits^w_0 f(y)dy}
\right]
\Psi(x)
=E\Psi(x),
\label{Eq_BdG_int6}
\end{equation}
The expression in the square brackets is the effective edge BdG Hamiltonian from the main text [cf. Eqs. (\ref{H}) and (\ref{Delta})].
It is worth reminding that this result holds for a generic local pair potential $\Delta_{_{2D}}(x,y)$.


\begin{thebibliography}{99}
\bibitem{Geometry}
The choice of the integration path depends on the screening properties of the superconductor. 
Here, it has the thin-film geometry in which the penetration 
of a perpendicular magnetic field is governed by the Pearl length. 
In practice, the Pearl penetration depth is comparable to or exceeds the half of the film length $L/2$ 
(see, e.g., Refs. \cite{Rohlfing09,Maier12,Oostinga13}).
In this case, the vertical branch of the integration path runs through the middle of the superconductor 
(at distance $L/2$ from the boundary in Fig. \ref{S_edge}) 
where $\partial_y \varphi =0$ for any symmetric magnetic field configuration.      
Besides, in order to avoid the corner points of the superconducting leads,
we take them to be wider than the half of the 2DTI, 
which, however, is not crucial for the physics.
\end{thebibliography}
\end{document}